\documentstyle[12pt]{article}
\begin{document}
\begin{center}
{\LARGE\bf{Reducing the Theoretical Uncertainty in Extracting
$|V_{ub}|$ from the Inclusive $B \rightarrow X_u\ell^- \overline{\nu}_\ell$ Decay
Rate}}\vspace{1cm}\\
{\large M. R. Ahmady$^*$, F. A. Chishtie$^*$, V. Elias$^*$, and T. G. Steele$^{\dagger}$}\vspace{1cm}\\
{\footnotesize\it $^*$Department of Applied Mathematics, The University of Western Ontario\\
London, Ontario  N6A 5B7  Canada.\\
$^{\dagger}$Department of Physics and Engineering Physics, University of Saskatchewan\\
Saskatoon, Saskatchewan  S7N 5E2  Canada.}
\end{center}
\vspace{-.9cm}
\bigskip
\bigskip
%\begin{abstract}
\noindent
{\footnotesize{\bf Abstract.} Utilizing asymptotic Pad\'{e}-approximant methods, we estimate the three-loop order 
$\overline{MS}$ coefficients of $\alpha_s^3[\log(\mu^2/m^2_b(\mu)]^k$ terms
$[k = \{0,1,2,3\}]$ within the $b \rightarrow u\ell^-\overline{\nu}_\ell$
decay rate. Except for the coefficient of the $k = 0$ term, all other
coefficients may also be obtained via renormalization-group (RG)
methods. The relative errors in asymptotic Pad\'{e}-approximant
estimates of the $k = \{1,2,3\}$ terms are found to be 5.1\% or less,
thereby providing an estimate of the theoretical uncertainty in the
asymptotic Pad\'{e}-approximant estimate of the $RG$-inaccessible $k =
0$ term. By a judicious choice in the renormalization scale parameter,
we are able to extract $|V_{ub}|$ from the inclusive decay rate to
within $\pm 7\%$ theoretical uncertainty.}
%\end{abstract}
\begin{center}
{\large\bf INTRODUCTION}
\end{center}
The inclusive semileptonic $B \rightarrow X_u \ell^-\overline{\nu}_\ell$ decay rate 
is sensitive only to a single mass parameter, the $b$-quark mass $m_b$, 
since the $u$-quark, charged lepton, and associated antineutrino 
are all very light compared to $m_b$. This property makes the calculated
decay rate particularly attractive for the extraction of the CKM
mixing parameter $|V_{ub}|$.  QCD corrections to the rate have recently 
been calculated to two-loop order within an $\overline{{\rm MS}}$
scheme [1]. If the renormalization 
scale $\mu$ is chosen to be $m_b(m_b) \cong 4.2\,GeV$ [2], 
i.e., if $\mu = m_b(\mu)$, then the two loop rate is given by 
\begin{equation}%1
\Gamma(B \rightarrow X_u\ell^-\overline{\nu}_\ell) = K m_b^5(m_b)\left[1 +
0.30 + 0.14\right], \; K \equiv G_F^2 |V_{ub}|^2/192\pi^3.
\end{equation}
Determinations of $|V_{ub}|$ based upon comparison of experimental results 
to this expression are necessarily subject to theoretical uncertainties 
associated with the premature truncation of the slowly convergent perturbative 
series in (1).  In the present work, we demonstrate how asymptotic Pad\'{e}-approximant 
methods in conjunction with renormalization-group (RG-) methods 
can dramatically reduce such theoretical uncertainties $\ldots$
\begin{enumerate}	
\item[1)] by providing an estimate of three loop contributions to the decay 
rate that can be tested against RG-accessible three-loop coefficients in the rate,  and 
\item[2)] by demonstrating that the physically-motivated 
choice of scale parameter $\mu$ which minimizes scale sensitivity [3] 
is quite close to that which minimizes the size of the four-loop
term, and consequently, the truncation error for an asymptotic series.
\end{enumerate}\vspace{.20cm}
\begin{center}
{\large\bf DETERMINATION OF RG-ACCESSIBLE THREE-LOOP COEFFICIENTS}\vspace{.5cm}
\end{center}

The series portion $S[x,L]$ of the perturbative $\overline{MS}$ expression for the 
inclusive semileptonic rate,
\begin{equation}%2
\Gamma\left(B \rightarrow X_u \ell^-\overline{\nu}_\ell\right) =
K\left[m_b^{(n_{f})} (\mu)\right]^5 S\left[x(\mu),L(\mu)\right],
\end{equation}
may be expressed in terms of the QCD expansion parameter
\begin{equation}%3
x(\mu) \equiv \alpha_s^{(n_{f})} (\mu)/\pi
\end{equation}
and n$^{th}$-order polynomials of logarithms
\begin{equation}%4
L(\mu) \equiv \log\left(\left[\mu/m_b^{(n_{f})}(\mu)\right]^2\right)
\end{equation}
appearing within the series coefficient of $x^n(\mu)$.  
The superscript $n_f$  indicates the number of active flavours 
contributing to the evolution of the QCD coupling constant $\alpha_s$ 
and the running $b$-quark mass. Thus, the series 
$S[x,L]$ within (2) is of the form
\begin{eqnarray}%5
S[x,L]   & = & 1 + x\left(a_0 + a_1 L\right) + x^2\left(b_0 + b_1 L + b_2
L^2\right)\nonumber \\
& + &  x^3\left(c_0 + c_1 L + c_2 L^2 + c_3 L^3\right) +
{\cal{O}}\left(x^4\right).
\end{eqnarray}
The one- and two-loop coefficients $a_0$, $a_1$, $b_0$, $b_1$, and $b_2$ 
have been calculated for arbitrary $n_f$ in [1].  For example, if $n_f  = 5$,
\begin{equation}%6
a_0 = 4.25360, \; a_1 = 5, \; b_0 = 26.7848, \; b_1 = 36.9902, \; b_2 = 17.2917.
\end{equation}
Note that the parameter $\mu$ within (3) and (4) is the (nonphysical) 
renormalization scale parameter.  The full decay rate (2) must ultimately be independent 
of the choice of $\mu$:
\begin{equation}%7
0  =  \frac{1}{Km_b^5} \left( \mu^2 \,\frac{d\Gamma}{d\mu^2}\right)
 =  \left[ (1 - 2\gamma(x)) \frac{\partial}{\partial L} + \beta(x) 
\frac{\partial}{\partial x} + 5\gamma(x) \right]S[x,L],
\end{equation}
\begin{equation}%8
\beta(x) = -\left(\beta_0 x^2 + \beta_1 x^3 + \beta_2 x^4 +
\ldots\right),
\end{equation}
\begin{equation}%9
\gamma(x) = -\left(\gamma_0 x +\gamma_1 x^2 + \gamma_2 x^3 +
\ldots\right),
\end{equation}
where, for $n_f = 5$, $\beta_0 = 23/12$, $\beta_1 = 29/12$, $\beta_2 =
9769/3456$, $\gamma_0 = 1$, $\gamma_1 = 253/72$, and $\gamma_2 = 7.4195$
[2].
If we expand the final line of (7) perturbatively in powers 
of $x^kL^m$, we find  that 
%10
$$0  =   x\left(a_1 - 5\gamma_0\right) + x^2\left(b_1 - a_0 \beta_0 -
5a_0 \gamma_0 + 2a_1 \gamma_0 - 5\gamma_1\right) +  x^2 L\left(2b_2 - a_0 \beta_0 - 5a_1 \gamma_0\right)\nonumber$$
$$ +  x^3 \left(c_1 - 2b_0 \beta_0 - a_0 \beta_1 - 5b_0\gamma_0 + 2b_1
\gamma_0 -  5a_0 \gamma_1 + 2a_1 \gamma_1 - 5\gamma_2\right)\nonumber$$
$$ +   x^3 L \left(2c_2 - 2b_1 \beta_0 - a_1\beta_1 - 5a_1 \gamma_1 -
5b_1 \gamma_0 + 4b_2 \gamma_0\right) \nonumber$$
\begin{equation}+ x^3L^2 \left(3c_3 - 2b_2 \beta_0 -  5b_2 \gamma_0\right) +
{\cal{O}}(x^4).
\end{equation}
For $n_f = 5$ we see from (6) that the coefficients 
of $x$, $x^2$, and $x^2 L$ vanish, as required.  
The requirement that $O(x^3)$ coefficients in (10) also vanish 
may be used to determine the logarithmic coefficients 
$c_1$, $c_2$, and $c_3$ within $S[x,L]$.  For $n_f = 5$, we find from (6) and (10) that
\begin{equation}%11
c_1 = 249.592,\;\; c_2 = 178.755,\;\; c_3 = 50.9144.
\end{equation}
The coefficient $c_0$, however, 
is RG-inaccessible to the order we are working, and must be obtained 
via a direct three-loop calculation\vspace{.20cm}.\\
\begin{center}
{\large\bf PAD\'E-APPROXIMANT ESTIMATION OF
NEXT-ORDER TERMS}
\end{center}

In the absence of a direct calculation, Pad\'{e}-approximant 
methods provide a means for estimating next-order terms in a perturbative series, 
provided such a series obeys appropriate criteria generally believed to be 
applicable to field-theoretical perturbative series [4].   As an example 
of the power of such methods, consider the following perturbative series:
\begin{equation}%12
S = 1 - x + \frac{1}{2} x^2 - \frac{1}{3} x^3 + \frac{5}{24} x^4 +
\ldots\;.
\end{equation}
This series is, in fact, the first five terms of the 
Maclaurin expansion of the function 
$S(x) = \sec(x) - \tan(x)$. A $[N|M]$ Pad\'{e} approximant to a series 
is the ratio of  degree-$N$ and degree-$M$ 
polynomials that replicates the known terms of a series.  
For example, the $[2|2]$ approximant which replicates the series (12) 
is specified (up to a common scale factor for the numerator and denominator) by 
\begin{equation}%13
S^{[2|2]} = \frac{1 - \frac{x}{2} - \frac{x^2}{12}}{1 + \frac{x}{2} -
\frac{x^2}{12}}.
\end{equation}
Specifically, the Maclaurin expansion of $S^{[2|2]}$ is
\begin{equation}%14
S^{[2|2]} = 1 - x + \frac{1}{2} x^2 - \frac{1}{3} x^3 + \frac{5}{24} x^4
- \frac{19}{144} x^5 + \ldots\; .
\end{equation}
The final series term in (14) may be regarded 
to be a Pad\'{e}-approximant prediction for the next term in the series (12). 
This prediction is surprisingly accurate, as the underlying function's Maclaurin series 
expansion is
\begin{equation}%15
\sec(x) - \tan(x) = 1 - x + \frac{1}{2} x^2 - \frac{1}{3} x^3 +
\frac{5}{24} x^4 - \frac{2}{15} x^5 + \ldots\; .
\end{equation}
We see that the final series terms listed in (14) and (15) are quite 
close: $19/144 (= 0.13194 \ldots) \cong  2/15 (= 0.13333\ldots)$.

In the calculation of the 
$B \rightarrow X_u \ell^-\overline{\nu}_\ell$ decay rate, the particular 
series with which we are concerned is of the form
\begin{equation}%16
S[x,L] = 1 + R_1[L]x + R_2[L]x^2 + R_3[L]x^3 + \ldots\;,
\end{equation}
with
\begin{equation}%17
R_1[L] = a_0 + a_1L,\;\;R_2[L] = b_0 + b_1L + b_2 L^2
\end{equation}
fully determined by the previously-calculated coefficients (6), 
and with the next order coefficient
\begin{equation}%18
R_3[L] = c_0 + c_1L + c_2L^2 + c_3 L^3
\end{equation}
assumed to be unknown. Of course, $c_1$, $c_2$, and $c_3$ are in fact 
accessible by RG methods [they are listed in (11)]; we will 
ultimately compare Pad\'{e}-estimates to RG-determinations of 
these coefficients to ascertain the accuracy of the Pad\'{e}-estimation 
procedure we utilise to obtain the RG-inaccessible constant $c_0$.

We seek a procedure to estimate $R_3$ in terms of the known lower order 
coefficients $R_1$ and $R_2$ appearing in (16). 
To do this, we assume that the error of an $[N-1|1]$ approximant in estimating 
the $N+1^{th}$  term in a series is inversely proportional to $N$.  
For example, the $[0|1]$ approximant to the series $1 + R_1x + R_2x^2 + R_3x^3 + \ldots$
is just
\begin{equation}%19
S^{[0|1]} = \frac{1}{1 - R_1 x} = 1 + R_1 x + R_1^2 x^2 + \ldots\;.
\end{equation}
The predicted coefficient of $x^2$ in (19) is 
$R_2^{[0|1]} \equiv R_1^2$, as opposed to the true 
coefficient $R_2$.  
We find the relative error of the $[0|1]$ approximant in predicting $R_2$ to be
\begin{equation}%20
\delta_{[0|1]} \equiv \frac{R_2^{[0|1]} - R_2}{R_2} = \frac{R_1^2 -
R_2}{R_2} .
\end{equation}
Similarly the $[1|1]$ approximant to the same series is
\begin{equation}%21
S^{[1|1]} = \frac{1 + (R_1 - R_2/R_1)x}{1-(R_2/R_1)x} = 1 + R_1 x +
R_2x^2 + \left(R_2^2/R_1\right)x^3 + \ldots\; .
\end{equation}
The predicted coefficient of $x^3$ is  $R_2^2/R_1$, 
which we denote as $R_3^{[1|1]}$, and the relative error 
associated with this prediction is 
\begin{equation}%22
\delta_{[1|1]} \equiv \frac{R_3^{[1|1]} - R_3}{R_3} = 
\frac{R_2^2/R_1-R_3}{R_3}.
\end{equation}
If the relative error of an $[N-1|1]$ approximant in predicting 
$R_{N+1}$ is inversely proportional to $N$, i.e., if
\begin{equation}%23
\delta_{[N-1|1]} \equiv \frac{R_{N+1}^{[N-1|1]} - R_{N+1}}{R_{N+1}} =
- \frac{A}{N},
\end{equation}
as suggested in [5], we see from (20) and (23) 
that $A = 1 - R_1^2/R_2$. Since (23) also implies 
that  $\delta_{[1|1]} = -A/2$, we can use (22) to obtain 
an error-improved estimate of $R_3$  entirely in terms of the 
known series terms $R_1$ and $R_2$ [6]:
\begin{equation}%24
R_3 = 2R_2^3/[R_1^3 + R_1R_2].
\end{equation}
\vspace{.20cm}
\begin{center}
{\large\bf ESTIMATION OF THREE-LOOP COEFFICIENTS}
\end{center}

The problem with the prediction (24) for the calculation considered here
is the incompatibility of $R_3$ as a degree-3 polynomial in $L$ (18)
with the rational (non-polynomial) function of $L$ one obtains via substitution of
(17) into (24). One approach to this problem is to generate a least
squares fit of (24) to the form (18) over the perturbative portion of
the domain of $L(\mu)$, i.e., over the ultraviolet domain $\mu >
m_b(\mu)$.  We define $w \equiv m_b^2 (\mu)/\mu^2$, in which case $L = -
\log(w)$ from (4), and obtain a least squares fit of (18) to (24) over
the range $0 \leq w \leq 1$ [i.e., $\mu \geq m_b(\mu)$] by optimizing
the function [7] 
$$\chi^2(c_0, c_1, c_2, c_3) \nonumber$$
\begin{equation}
 \equiv \int_0^1 dw\left[
\frac{2R_2^3(w)}{R_1^3(w) + R_1(w)R_2(w)}
- \left(c_0 - c_1 \log (w) + c_2 \log^2 (w) - c_3 \log^3
(w)\right)\right]^2
\end{equation}
with respect to $c_i$, the polynomial coefficients of powers of $L$
within (18). Upon substitution of the known [via (6) and (17)]
functions $R_1(w) = a_0 - a_1 \log(w)$, $R_2(w) = b_0 - b_1 \log(w) +
b_2 log^2 (w)$ into the integrand of (25), we find from explicit
numerical integration that
$$\chi^2(c_0, c_1, c_2, c_3) = c_0^2 + 2c_1^2 + 24c_2^2 
+ 720c_3^2 + 2c_0c_1 + 4c_0c_2 + 12c_0c_3\nonumber$$
$$ +  12c_1c_2 + 48c_1c_3 + 240c_2c_3
 -  2237.27c_0 - 5980.96c_1\nonumber $$
\begin{equation}
 -  24423.1c_2 - 129077c_3 +  6386542 .
\end{equation}
The four conditions $\partial\chi^2/\partial c_i = 0$ ($i = \{0, 1, 2,
3\}$) provide the following Pad\'{e}-estimates for $c_i$:
\begin{equation}
c_0 = 198, \; c_1 = 261, \; c_2 = 184, \; c_3 = 48.6\; .
\end{equation}
Note the excellent agreement of estimates for $\{c_1, c_2, c_3\}$ with the
true values (11) obtained via RG methods. The relative errors $\delta
c_i \equiv (c_i - c_i^{RG})/c_i^{RG}$ in the Pad\'{e} estimates are
respectively $\delta c_1 = -4.6\%$, $\delta c_2 = 2.9\%$, $\delta c_3 =
2.1\%$. These small errors suggest similar accuracy in the determination
of the unknown coefficient $c_0$. Utilizing this value of $c_0$ in the
$L = 0$ limit, appropriate for the $\mu = m_b(\mu)$ choice of
renormalization scale in (1), we find that the series within (1) is now improved to
\begin{equation}
\Gamma\left(B \rightarrow X_u \ell^- \overline{\nu}_\ell\right)  = 
K(4.17\,GeV)^5[1 + 0.304 + 0.137 + 0.075] =  K(1913\, GeV^5).
\end{equation}
The estimate (28) is obtained using the central value of Chetyrkin and
Steinhauser's [8] determination of $m_b(m_b) = 4.17 \pm 0.05\,GeV$ an
estimate which followed from a central value $\alpha_s(M_z) = 0.118$
[also utilized to evolve $x(\mu)$ in (28)]\vspace{.20cm}.\\
\begin{center}
{\large\bf RESIDUAL SCALE DEPENDENCE}
\end{center}

The series within the rate (28) is evaluated at $\mu = 4.17\,GeV$ and
appears to be the first four terms of a positive-term perturbative
series. This positive-term character is evident for $\mu > 4.17\,GeV$ as
well, the domain of $L(\mu)$ for which $L$ is positive, since the
coefficients  $a_i$, $b_i$, and $c_i$ (including the estimated value for
$c_0$) are all positive. If convergent, a truncated positive-term series
necessarily underestimates the series sum. For values of $\mu$ larger
than $4.17\,GeV$, this underestimation is even more severe, because of the
positive growth of $L(\mu)$. Thus $\Gamma(\mu = 9\,GeV) = K(1733\,GeV^5) <
\Gamma(\mu = 4.17\,GeV)$, as evident from (28).

As noted in [1], the perturbative series
does not become locally flat until $\mu$ is well into the infrared 
region.   This region $[\mu < m_b(m_b)]$ 
can be accessed via threshold matching conditions [9] to a running
coupling constant and running mass characterized by four active
flavours:
%29
$$x^{(4)}\left[m_b(m_b)\right]\nonumber$$
\begin{equation}
 =   x^{(5)}\left[m_b(m_b)\right]
\left[1  +   0.1528\left(x^{(5)}\left[m_b(m_b)\right]
\right)^2
+ 0.633\left(x^{(5)}\left[m_b(m_b)\right]\right)^3\right],
\end{equation}
%30
$$m^{(4)}_b\left[m_b(m_b)\right]\nonumber$$
\begin{equation}
 =   m^{(5)}_b\left[m_b(m_b)\right]
\left[1  +   0.2060\left(x^{(5)}\left[m_b(m_b)\right]
\right)^2
+ 1.95\left(x^{(5)}\left[m_b(m_b)\right]\right)^3\right].
\end{equation}
If $\alpha_s(M_z) = 0.118$ and $m_b^{(5)} (m_b) = 4.170\,GeV$, the
conditions (29) and (30) imply that $x^{(5)}(4.170\,GeV) = 0.07155$,
$x^{(4)}(4.170\,GeV) = 0.07162$, and $m_b^{(4)}(4.170\,GeV) = 4.177\,GeV$.
Moreover, with 4 active flavours the two-loop constants characterising
the rate (5) are no longer as given in (6); instead one finds for $n_f =
4$ that [1] 
\begin{equation}
b_0 = 25.7547,\; b_1 = 38.3935,\; b_2 = 17.7083,
\end{equation}
and, from perturbative RG-invariance (10) with constants $\beta_i$ and
$\gamma_i$ characterized by four active flavours, 
\begin{equation}
c_1 = 263.84,\; c_2 = 194.23,\; c_3 = 54.109.
\end{equation}
One can also obtain estimates of the $n_f = 4$ three-loop coefficients
via optimization of the $\chi^2$ function (25):
\begin{equation}
c_0 = 181.5,\; c_1 = 277.3,\; c_2 = 197.6,\; c_3 = 51.86.
\end{equation}
The startlingly close agreement between (32) and (33) for the RG-
accessible coefficients $c_1$, $c_2$, $c_3$ is indicative of comparable
accuracy for
the estimate of $c_0$ in (33). 

In ref. [10], it is shown that the $n_f = 4$ values for $c_i$ are
consistent with continuity of the rate $\Gamma(B \rightarrow X_u \ell^-
\overline{\nu}_\ell$) across the $n_f = 5$ threshold at $4.17\,GeV$. The rate
is shown to increase with decreasing $\mu$ until a local maximum is reached at
$\mu = 1.775\,GeV$[10]:
\begin{equation}
\left(\frac{d\Gamma}{d\mu}\mid_{\mu = 1.775\,GeV}\right) = 0,
\end{equation}
%35
$$\Gamma\left[ \mu  =  1775\,GeV\right]\nonumber$$
\begin{equation}
 =  K\left[m_b^{(4)} (1.775\,GeV)\right]^5\left[1 - 0.6455+ 
0.2477 - 0.0143 \right] = K(2071\,GeV^5).
\end{equation}
The optimisation condition (34) ensures that $\mu = 1.775\,GeV$ is the choice of $\mu$
for which there is minimal residual scale sensitivity of the rate, an
established criterion for the choice of scale [3]. Note that the
perturbative series within (35) is now an {\it{alternating series}}
whose 3-loop term is much smaller than that in (28). The alternation of
sign becomes possible because $L(\mu)$ is negative when $\mu <
m_b(\mu)$, as evident from (4). If the series within (35) continues to
be both alternating and montonically decreasing, the rate at $\mu =
1.775\,GeV$ can be shown to be bounded from below {\it and} above:
\begin{equation}
K \times 2071\,GeV^5 \leq \Gamma\left[\mu = 1.775\,GeV\right] \leq K \times 2122\,GeV^5.
\end{equation}

Remarkably, the ``minimal-sensitivity'' point defined by (34) is quite
close to the point at which the three-loop term vanishes entirely, a
point at which three-loop estimation of
the series, if asymptotic, would be most accurate.
With RG-values (32) for $c_1$, $c_2$, $c_3$ and the Pad\'{e}-
estimate $c_0 = 188.5$ for $c_0$ [corresponding to minimization of
$\chi^2$ with respect to $c_0$ after explicit incorporation of RG-values
of $c_1$, $c_2$, $c_3$ into (25)], the three loop term (18) is seen to
{\it{vanish}} at $\mu = 1.835\,GeV$, at which point $\Gamma(\mu =
1.835\,GeV) = K \times 2069\,GeV^5$. The agreement between this value for
the rate with the minimal-sensitivity estimate (35) is striking\vspace{.20cm}.
\begin{center}
{\large\bf $|V_{ub}|$: THEORETICAL UNCERTAINTIES}
\end{center}

The minimal sensitivity rate (35) is subject to the following $\Delta \Gamma/K$
theoretical uncertainties: truncation error (t.e.)
\begin{equation}%37
|\Delta\Gamma/K|_{\rm{t.e.}} = \left[m_b^{(4)} (1.775)\right]^5 (0.0143)
= 51\,GeV^5,\end{equation}
error in estimating $c_0 \left(|\Delta c_0/c_0| \simeq |\Delta c_i/c_i^{RG}|
\simeq 5\% \right)$,
\begin{equation}%38
|\Delta\Gamma/K|_{c_{0}} = 35\,GeV^5,
\end{equation}
uncertainty in $\alpha_s(M_z) (= 0.119 \pm 0.002$ [2]),
\begin{equation}%39
\Delta\Gamma/K|_{\alpha_{s}} = \left(
\begin{array}{c}
+108\\
-28\end{array}
\right) GeV^5
\end{equation}
(asymmetric bounds are because (35) is based upon $\alpha_s(M_z) =$
 0.118), uncertainty in $m_b(m_b) (= 4.17 \pm 0.05$ [8]),
\begin{equation}%40
\Delta\Gamma/K|_{m_{b}} = \left(\begin{array}{c}
+120 \\
-115\end{array}\right)
GeV^5,\end{equation}
and the asymmetric nonperturbative (n.p) contribution to the overall rate [11]:
\begin{equation}%41
\Delta\Gamma/K|_{\rm{n.p.}} = \left(\begin{array}{c}
-39 \\
-57\end{array}\right)
GeV^5.\end{equation}
Inclusive of all these contributions, the result (35) becomes
\begin{equation}%42
\Gamma = K \times (2065 \pm 14\%)GeV^5.
\end{equation}
Using (1) for the coefficient $K$, the branching ratio for inclusive
charmless semileptonic decay is seen to be
\begin{equation}%43
R \equiv \Gamma\left( B \rightarrow X_u \ell^-
\overline{\nu}_\ell\right)/\Gamma(B \rightarrow {\rm{anything}})
= \frac{|V_{ub}|^2G_F^2}{192\pi^3} \,\frac{\left[(2065 \pm
14\%)GeV^5\right]}{4.25 \cdot 10^{-13} GeV},
\end{equation}
in which case
\begin{equation}%44
|V_{ub}| = (0.0949 \pm 0.0066)R^{1/2}.\end{equation}
The central value of (44) is in agreement with Uraltsev's relation between
$|V_{ub}|$ and the branching ratio [12], and is indicative of an overall
$\pm 7\%$ theoretical uncertainty in this relation.
\eject
\begin{center}
{\large\bf{REFERENCES}}
\end{center}
{\footnotesize{
1. van Ritbergen, T., {\it Phys. Lett. B} {\bf 454}, 353 (1999).\\
2. Caso, C. et al. (Particle Data Group), {\it Eur. Phys. J. C} {\bf 3},
1 (1998).\\
3. Stevenson, P.M., {\it Phys. Rev. D} {\bf 23}, 2916
(1981).\\
4. Samuel, M. A., Li, G., and Steinfelds, E., {\it Phys. Rev. E} {\bf 51},
3911 (1995); Samuel, M. A. and Druger, S. D., {\it Int. J. Th. Phys.} {\bf 34},
 903 (1995).\\
5. Ellis, J., Karliner, M., and Samuel, M. A., {\it Phys. Lett. B} {\bf
400}, 176 (1997).\\
6. Elias, V., Steele, T. G., Chishtie, F., Migneron, R., and Sprague,
K., {\it Phys. Rev. D} {\bf 58}, 116007 (1998).\\
7. Chishtie, F.A., Elias, V., and Steele, T. G., {\it J. Phys. G} {\bf
26}, 93 (2000).\\
8. Chetyrkin, K. G. and Steinhauser, M., {\it Phys. Rev. Lett.} {\bf
83}, 4001 (1999).\\
9. Chetyrkin, K. G., Kniehl, B. A., and Steinhauser, M., {\it Nucl.
Phys. B} {\bf 510}, 61 (1998).\\
10. Ahmady, M. R., Chishtie, F. A., Elias, V., and Steele, T. G., {\it
Phys. Lett. B} {\bf 479}, 201 (2000).\\
11. Hoang, A. H., {\it Nucl. Phys. Proc. Suppl.} {\bf 86}, 512 (2000);
Bagan, E., Ball, P., Braun, V.M.,
Gosdzinsky, P., {\it Phys. Lett. B} {\bf 342}, 362 (1995).\\
12. Uraltsev, N., {\it Int. J. Mod. Phys. A} {\bf 11}, 515 (1996).}}\\
\end{document}